\documentclass{epl}

\newcommand{\PR}{PrOs$_4$Sb$_{12}$~}
\newcommand{\vq}{\bf q\rm~}
\newcommand{\vk}{{\bf k\rm}}
\newcommand{\vH}{\bf H\rm~}
\newcommand{\vv}{\bf v\rm~}
\newcommand{\tom}{$\tilde{\omega}_n$}
\newcommand{\tde}{$\tilde{\Delta}$}

\newcommand{\la}{\langle}
\newcommand{\ra}{\rangle}
\newcommand{\De}{$\Delta(\bf k\rm)~$}

\title{Multiphase Superconductivity in Skutterudite PrOs$_4$Sb$_{12}$}

\shorttitle{}

\author{K.~Maki\inst{1}\thanks{E-mail: \email{kmaki@usc.edu}} \and
H. Won\inst{2} \and P.~Thalmeier\inst{3} \and Q. Yuan \inst{3,4} \and
K. Izawa\inst{5} \and Y. Matsuda\inst{5}}

\shortauthor{K.~Maki \etal}

\institute{
\inst{1} Dept. of Physics and Astronomy, University of Southern California --\\
Los Angeles, CA 90089-0484 U.S.A.\\
\inst{2} Dept. of Physics, Hallym University Chuncheon 200-702, South Korea\\
\inst{3} Max Planck Institute for the Chemical Physics of Solids --
Dresden, Germany\\
\inst{4} Pohl Institute of Solid State Physics, Tongji University -- 
Shanghai 200092, P.R.China\\
\inst{5} Institute for Solid State Physics, University of Tokyo --
Kashiwanoha 5-1-5, Kashiwa, Chiba 277-8581, Japan}




\makeatletter
\def\ps@epl@titlepage{%
   \def\@oddfoot{}%
   \let\@evenfoot\@empty%
   \let\@evenhead\@empty%
   \def\@oddhead{}%
   \let\@mkboth\@gobbletwo
   \let\sectionmark\@gobble
   \let\subsectionmark\@gobble
}
\def\epl@shortbanner{\null}
\makeatother

\begin{document}

\maketitle

\begin{abstract}
We propose a model of nodal order parameters for the superconducting A
and B - phase of 
skutterudite PrOs$_4$Sb$_{12}$ and discuss the associated angular dependent
magnetothermal conductivity at low temperatures. The model for the
hybrid gap functions \De containing an s- and g- wave parts is consistent
with recent thermal conductivity experiments in PrOs$_4$Sb$_{12}$. In
particular the model accounts for the data on polar and
azimuthal field angle dependence of $\kappa_{zz}(\theta,\phi)$. The
low temperature behaviour of thermodynamic properties in zero field is also
presented. We show that the effect of impurity scattering on a nodal
hybrid gap function immediately leads the opening of a gap and
related exponential behaviour of low temperature specific heat and
thermal conductivity which is very different from d- wave superconductors.

\end{abstract}

\section{Introduction}

Superconductivity in the cubic (T$_h$ symmetry) Heavy Fermion (HF)
skutterudite PrOs$_4$Sb$_{12}$ with
$T_c$=1.8 K has provoked great interest since it exhibits a number of
characteristics that suggest the presence of
nodes.\cite{Bauer02,Vollmer02,Kotegawa02} However
there is also a proposal that it is a conventional s-wave
superconductor.\cite{MacLaughlin02} The specific heat
results\cite{Bauer02,Vollmer02} indicate
that firstly a low temperature power law behaviour and secondly the
presence of two superconducting (sc) phases with two consecutive specific
heat jumps at $T_{c1}$=1.82 K and $T_{c2}$=1.75 K somewhat similar to the
observations in UPt$_3$. More
recent low temperature thermal conductivity results\cite{Izawa02a} in
the vortex state confirm i) the presence of two superconducting A
($T<T_{c1}$) and B ($T<T_{c2}$) phases with
different nodal structure, ii) the presence of point nodes in both A
($H>0.75$T, $T\ll T_c$) and B ($H<0.75$T, $T\ll T_c$) phase. More
specifically the thermal
conductivity indicates the presence of four point nodes at \vk=
(1,0,0), (0,1,0), (-1,0,0) and (0,-1,0) (in units of k$_F$) in \De in
the A-phase while in the B-phase the two point nodes are located at \vk=
(0,1,0) and (0,-1,0). Present investigation also excludes a fully symmetric A-
phase order parameter with additional point nodes along [001]
direction.

\section{Symmetry classification of SC order parameter}

The presence of two sc phases is possibly connected with the vicinity of
a field induced antiferroquadrupolar ordered phase above 4T which
is likely due to a level crossing of the tetrahedral CEF
states \cite{Takegahara01} where one of the excited triplet $\Gamma_5$
states crosses the singlet ground state $\Gamma_1$. In U-based
unconventional HF
superconductors an exchange of (dipolar) spin fluctuations in the itinerant 
5f-electrons is frequently implied as the mechanism for unconventional SC
pair formation. In the present case a new interesting possibility for pair
formation is the exchange of inelastic quadrupolar $\Gamma_1$-
$\Gamma_5$ fluctuations of the
essentially localized 4f electrons of Pr. Since the AFQ
critical field is of the same order of magnitude as the sc upper
critical field of H$_{c2}$(0) = 2T impeding CEF level crossing as
function of applied mangetic field will
strongly affect the pair potential and may lead to a change of the sc
gap symmetry. This does however not explain the origin of the zero
field T$_c$ splitting. Details of
this microscopic mechanism and its prefered gap symmetries are not
clear, hence we have to consider possible order parameters compatible
with the above node structure on a phenomenological basis. The gap
function may be expanded in terms of basis functions $\psi^\Gamma_i(\vk)$
which transform like representations $\Gamma$ of the crystal symmetry
group (i=1-d is the degeneracy index, the index l denoting the degree
of $\Gamma$ is suppressed). So far there is no information from NMR
Knight shift or H$_{c2}$- Pauli limiting whether PrOs$_4$Sb$_{12}$ has
spin singlet or triplet pairing, although recent $\mu$SR
measurements\cite{Aoki03} seem to suggest the presence of condensate
magnetic moments. Only the singlet case will be considered here. The
gap function should then be given by

\begin{eqnarray}
\label{EXPAND}
\Delta(\vk)&=&\sum_{\Gamma,i}
\eta^\Gamma_i\psi^\Gamma_i(\vk)\equiv\Delta f(\vk)
\end{eqnarray}

where the form factor $f(\vk)$ is
normalized to one and $\Delta$ is the temperature dependent maximum gap
value. In the spirit of the Landau theory only a {\em single}
representation with the highest $T_c$ should be realized and for $T\geq T_c$, 
the free energy may then be expanded in terms of possible invariants 
of the order parameter components\cite{Volovik85} $\eta^\Gamma_i$ 
which are determined by Landau parameters $\alpha^\Gamma (T)$ and
$\beta^\Gamma_i$. The node structure is then fixed by the specific
symmetry class of $\Delta$(\vk) defined by the set of $\eta^\Gamma_i$.  
However we are interested in the complementary temperature range
$T\ll T_c$ relevant for magnetothermal conductivity experiments. In
this regime Landau expansion is unreliable. Especially
it is unjustified to assume that only a {\em single}
representation $\Gamma$ will have appreciable amplitude in
Eq.~(\ref{EXPAND}), one should expect that at low temperatures 
$\Delta$(\vk) will be a hybrid gap function, i.e. a superposition 
of basis functions belonging to a few favorable representations 
which have approximately equal $T^\Gamma_c$. A striking realization 
of such a hybrid gap function has recently been found in the 
nonmagnetic borocarbide superconductor YNi$_2$B$_2$C\cite{Izawa02b} 
and possibly also in LuNi$_2$B$_2$C. There
\De=$\Delta f(\vk)$ is a sum of two fully symmetric (for
D$_{4h}$) gap functions of different degree (`s+g-wave') given by

\begin{eqnarray}
\label{GAP1}
f(\vk)&=&\frac{1}{2}[1-(k_x^4+k_y^4-6k_x^2k_y^2)]
\end{eqnarray}

For hybrid gap functions nodes occur only for special
`fine tuning' of amplitudes, they are not enforced by the gap
symmetry.\cite{Maki02,Thalmeier02}. Microscopic model calculations\cite{Yuan02}
show that this is possible for a wide variety of plausible pairing
potentials. The case of a nodal hybrid order parameter is realized in
YNi$_2$B$_2$C to an astonishing accuracy. This is not yet clear for
PrOs$_4$Sb$_{12}$ but the nodal structure refered above cannot be
realized with gap functions corresponding only to a single
T$_h$-representation. Experimental evidence from the field angle
dependence of $\kappa_{zz}(\theta,\phi)$ leads us to the
following simple proposals for hybrid gap functions \De
=$\Delta f(\vk)$ for PrOs$_4$Sb$_{12}$ with four and two nodal points
respectively:

\begin{eqnarray}
\label{GAP2}
\mbox{A-phase:~} f(\vk)&=&1-k_x^4-k_y^4, \qquad
\mbox{B-phase:~} f(\vk)=1-k_y^4.
\end{eqnarray}

Both A- phase and B-phase are described by hybrid gap functions which
are superpositions of three T$_h$ representations. The B-phase
gap function has lower symmetry and both are threefold 
degenerate. An alternative gap function for the A-phase of 
PrOs$_4$Sb$_{12}$ would be  the same model of eq.~(\ref{GAP1}) as used
already for the borocarbides.

\section{Quasiparticle DOS and thermodynamics}

\begin{figure}
\twofigures[width=60mm]{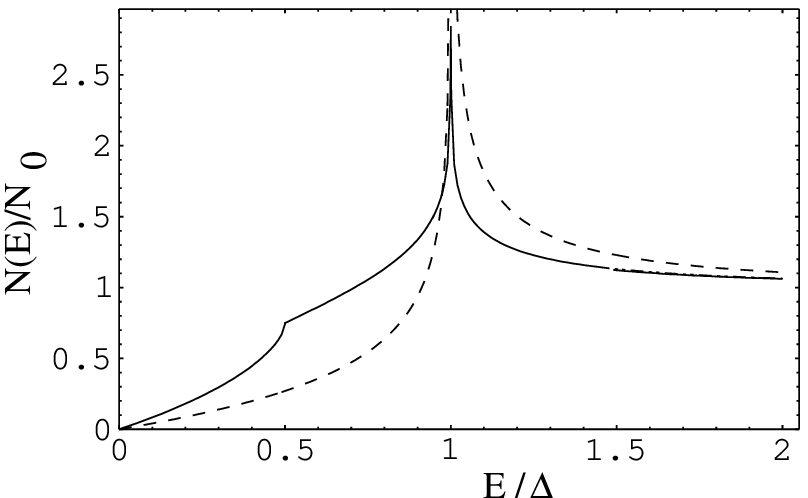}{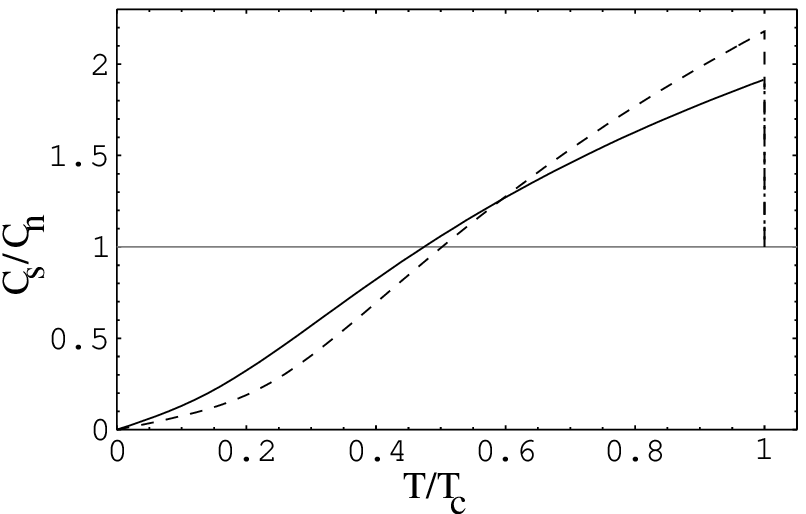}
\caption{Quasiparticle DOS for the A- phase gap functions (full line)
and B- phase gap function (dashed line).}
\label{FIGDOS}
\caption{Temperature dependence of specific heat $C_s$ ($C_n=\gamma_n T$)
for the A- phase gap functions (full line) and B- phase gap
function (dashed line). The specific heat jumps $\Delta C_s/C_n$ at
$T=T_c$ are given by 0.92 and 1.18, respectively.}
\label{FIGTEMP}
\end{figure}

In the normal state the estimated linear specific heat coefficient 
is\cite{Bauer02} $\gamma\geq$350 mJ/mole K$^2$, considerably larger than
the value obtained from the effective dHvA masses (m$^*\sim$2.4-7.6
m).\cite{Sugawara02} However in comparison with other Pr-compounds
both experiments support the HF character of PrOs$_4$Sb$_{12}$.
We have calculated the quasiparticle DOS N(E) for the gap functions of
A and B phase which determines the low temperature
thermodynamics. The DOS functions are shown in Fig.~\ref{FIGDOS}. For
E/$\Delta\ll$1 the lowest linear term N(E)/N$_0$=$\alpha$(E/$\Delta$) has
$\alpha= \frac{\pi}{4}$ for A phase  and $\frac{\pi}{8}$ for B phase.
Here $N_0$ is the normal state DOS. 
Correspondingly, the low $T$ specific heat, spin susceptibility and
superfluid density are given by, respectively,

\begin{eqnarray}
\label{THERMO}
\frac{C_s}{\gamma_nT}&=&
\alpha\frac{27}{\pi^2}\zeta(3)(\frac{T}{\Delta}),\qquad
\frac{\chi_s}{\chi_n}=\alpha(2\ln 2)\frac{T}{\Delta}\nonumber\\
\mbox{A-phase:~}\qquad
\hat{\rho}^{ab}_s(T)&=&1-\frac{3}{2}\frac{\chi_s}{\chi_n}\qquad
\hat{\rho}^c_s(T)=1-c_A\bigl(\frac{T}{\Delta}\bigr)^2\\
\mbox{B-phase:~}\qquad
\hat{\rho}^b_s(T)&=&1-3\frac{\chi_s}{\chi_n}\qquad
\hat{\rho}^a_s(T)=\hat{\rho}^c_s(T)=1-c_B(\frac{T}{\Delta}\bigr)^2\nonumber
\end{eqnarray}

Here $\hat{\rho}^i_s(T)\equiv\rho^i_s(T)/\rho^i_s(0)$ for the axis
directions i=a,b,c and c$_{A,B}$ are numerical constants.
Furthermore the specific heat in the whole temperature region is calculated 
and given by Fig.~\ref{FIGTEMP}, where we have assumed
phenomenologically the pairing
interaction $V_{\bf kk'}$ to be of the separable form $Vf({\bf
k})f({\bf k'})$. 
The $C_s/(\gamma_n T)$ vs. $T/T_c$ is found to be nearly independent of $V$ 
for each gap model. However for fixed V the obtained T$_c$ values for
A and B phase are quite different.
We consider now thermodynamics in the vortex state and in the low
temperature limit with $\Gamma\ll T\ll v\sqrt{eH}\ll\Delta(0)$ where
where $v$ is the (isotropic) Fermi velocity and $\Gamma$ the quasiparticle
scattering rate and $v\sqrt{eH}$ is the typical magnetic energy. 
The field $\vH$ is applied in an
arbitrary direction defined by polar and azimuthal angles $\theta$ and
$\phi$ respectively refered to a cubic [001] axis. In \cite{Volovik93,Won01}
it has been established that the dominant effect of the
magnetic field in nodal superconductors is the Doppler shift in the
quasiparticle spectrum introduced by the supercurrent around each
vortex. A short derivation is also sketched in Ref.~\cite{Maki02}. 
In the low T limit the quasiparticle DOS for E=0 in the two phases
(i=A,B) is given by

\begin{eqnarray}
\label{DOPP}
\frac{N(0)}{N_0}&=&\frac{\pi}{4}\la|x|\ra =g(H,\theta,\phi), \;\;\;\;
\la|x|\ra_i= \frac{2}{\pi}\frac{v\sqrt{eH}}{\Delta}I_i(\theta,\phi).
\end{eqnarray}

Here $x={\bf v} \cdot {\bf q}/\Delta$ is the normalized Doppler shift
energy and the averaging over the Fermi surface and vortex lattice has
to be performed. For brevity we use $x_A\equiv \la|x|\ra_A$ and  
$x_B \equiv \la|x|\ra_B$. The angular dependent functions
I$_{A,B}(\theta,\phi$) are given by 

\begin{eqnarray}
I_A(\theta,\phi)&=&\frac{1}{2}
[(1-\sin^2\theta\sin^2\phi)^\frac{1}{2}
+(1-\sin^2\theta\cos^2\phi)^\frac{1}{2}]\ ,\nonumber\\
I_{B}(\theta,\phi)&=&\frac{1}{4}(1-\sin^2\theta\sin^2\phi)^\frac{1}{2}]
\end{eqnarray}

This leads to a field angle dependent specific heat for the A and B
phase which is given by

\begin{eqnarray}
C_{s}/\gamma_nT=\frac{v\sqrt{eH}}{2\Delta}I_i(\theta,\phi)\ .
\end{eqnarray}

Similar angular dependences governed by I$_{A,B}(\theta,\phi)$ may
be obtained for the spin susceptibility $\chi_s$(T) and superfluid density
$\rho_s$(T). In principle the angular dependent specific heat is adequate
to identify the point nodes in \De. Recently a related experiment is
reported for YNi$_2$B$_2$C.\cite{Park02} The observed cusps in the
angular dependence of $C_s$ clearly indicate the presence of point
nodes along the tetragonal plane axes at (1,0,0), (0,1,0) etc., which
is fully consistent with thermal conductivity data \cite{Izawa02b}.

\section{Impurity scattering and magnetothermal conductivity}

In order to compute the thermal conductivity, it is imperative to
consider the effect of impurity scattering. As already discussed
elsewhere the impurity scattering in ``s+g''- wave
supercondutors is completely different from other nodal
superconductors without s- wave component\cite{Yuan03,Won03}. Since
the unitary limit gives practically the same result as Born limit, we
can follow Abrikosov and Gor'kov\cite{Abrikosov61} and we have

\begin{eqnarray}
\label{AG}
\tilde{\omega}_n &=& \omega_n + \Gamma \big< \frac{\tilde{\omega}_n}
{\sqrt{\tilde{\omega}_n^2 + (\tilde{\Delta}+\Delta f')^2}}\big>
\nonumber\\
\tilde{\Delta}&=& a\Delta +\Gamma \big\langle
\frac{(\tilde{\Delta}+\Delta f')}{\sqrt{\tilde{\omega}_n^2 + 
(\tilde{\Delta}+\Delta f')^2}}\big\rangle
\end{eqnarray}

where \tom~ and \tde~ are renormalized Matsubara frequency and sc order
parameter respectively. Here a = $\frac{3}{5}$ and f'=
$\frac{2}{5}$-k$_x^4$-k$_y^4$ for A phase while a = $\frac{4}{5}$ and 
f'=$\frac{1}{5}$-k$_y^4$ for the B phase. First we consider H=0. In
the limit of small $\Gamma$ it is shown that \tde = a$\Delta$+$\Gamma$
which implies that a quasiparticle energy gap opens up
immediately. These energy gaps are well approximated by $\omega_g \simeq
\Gamma(1+\frac{\Gamma}{a\Delta})^{-1}$. This leads to
C$_s$/$\gamma_n$T and $\kappa_{ii}$/T $\sim
\bigl(\frac{\omega_g}{T}\bigr)^\frac{3}{2} e^{-\omega_g/T}$.
Therefore unlike d- wave superconductors there will be no universal
heat conduction \cite{Lee93,Sun95}. The thermal conductivity vanishes
exponentially when $T/\omega_g\ll1$. 

In the presence of a magnetic field \tom~ in r.h.s of eq.~(\ref{AG})
have to be replaced by \tom-i\vv$\cdot$\vq where \vv$\cdot$\vq is the
Doppler shift. then for $\frac{\Gamma}{\Delta}\ll$1 the quasiparticle
damping C$_0$=$\lim_{\omega\rightarrow
0}\frac{\tilde{\omega}_n}{\Delta}$ in A and B phase is given by

\begin{eqnarray}
C^i_0=\frac{\Gamma}{b_i\Delta}x_i\ln\bigl(\frac{2}{x_i}\bigr)
(1+\frac{31}{64}x_i^2 ..)
\end{eqnarray}

where x$_i$ (i=A,B) are defined below eq.~(\ref{DOPP}) and b$_A$ = 2,
b$_B$ = 4. A similar expansion in terms of x$_i$ up to second order is used
to obtain the thermal conductivity tensor:

\begin{eqnarray}
\label{AKAP}
\frac{\kappa^A_{zz}}{\kappa_n}&=&\frac{x_A}{\ln\bigl(\frac{2}{x_A}\bigr)}
\frac{1+\frac{\pi}{12}x_A+\frac{31}{40}x_A^2}{1+\frac{31}{64}x_A^2}\nonumber\\
\frac{\kappa^A_{xx}}{\kappa_n}&=&
\frac{3}{2\ln\bigl(\frac{2}{x_A}\bigr)}\bigl(\frac{x'}{x_A}\bigr)^2
\frac{1+\frac{1}{3\pi}x'+\frac{5}{128}x'^2}{1+\frac{31}{64}x_A^2}
\end{eqnarray}

where we have defined 
$x'\equiv\frac{2}{\pi}\frac{v\sqrt{eH}}
{\Delta}(1-\sin^2\theta\cos^2\phi)^\frac{1}{2}$. The component
$\kappa^A_{yy}$ may be obtained by replacing $x'\rightarrow x_B$ in
$\kappa^A_{xx}$. Similarly for the B- phase we obtain 

\begin{eqnarray}
\label{BKAP}
\frac{\kappa^B_{zz}}{\kappa_n}&=&\frac{\kappa^B_{xx}}{\kappa_n}=
\frac{x_B}{\ln\bigl(\frac{2}{x_B}\bigr)}
\frac{1+\frac{\pi}{12}x_B+\frac{31}{40}x_B^2}{1+\frac{31}{64}x_B^2}\nonumber\\
\frac{\kappa^B_{yy}}{\kappa_n}&=&
\frac{3}{2\ln\bigl(\frac{2}{x_B}\bigr)}
\frac{1+\frac{1}{3\pi}x_B+\frac{5}{128}x_B^2}{1+\frac{31}{64}x_B^2}
\end{eqnarray}

Note that in $\kappa^A_{xx}$ and $\kappa^B_{yy}$ have no linear term
in x$_i$ and therefore don't exhibit a $\sqrt{H}$ behaviour.
Indeed eqs.~(\ref{AKAP},\ref{BKAP}) reproduce the $\phi$ dependence
of $\kappa^{A,B}_{zz}$ observed in \cite{Izawa02a}. For $\theta=\pi/2$  
cusps in $\kappa_{zz}$ at $\phi=0,\ \pi/2$ (A phase) and
$\phi=\pi/2$ (B phase) are predicted in accordance with experiment.
On the other hand the present theory
predicts the $\sqrt{H}$ dependence of $\kappa_{zz}$ while the data
appear to be approximately linear in H. Also when the heat current 
is parallel to a pair of nodal directions the dominant term is almost
independent of H except for the $\ln\bigl(\frac{2}{x}\bigr)$ term. Also we have
neglected the $\Gamma$ dependent terms completely due to
T$\gg\Gamma$. However, in a more realistic situtation the inclusion of
the scattering term is necessary.

\begin{figure}
\onefigure[width=60mm]{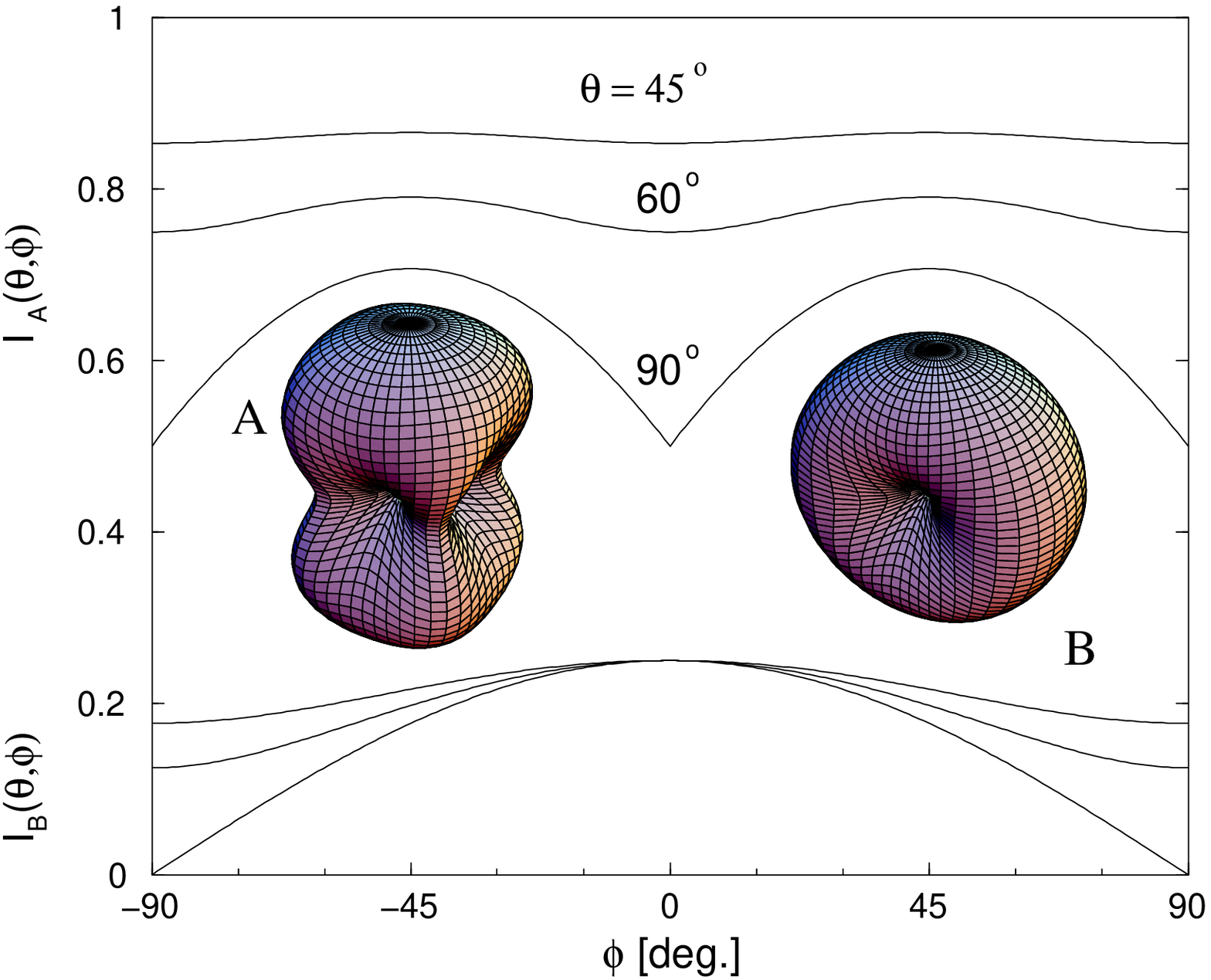}
\caption{Comparison of angular functions I$_{A,B}(\theta,\phi)$ for
the fourfold (A) and twofold (B) gap functions. Inset show polar plots
of the two gap functions with nodes along [100] and [010] directions
for A- phase and along [010] for B- phase.}
\label{FIGANGI}
\end{figure}

In Fig.~\ref{FIGANGI} we show I$_{A,B}$($\theta,\phi$) 
which determines the $\theta,\phi$- dependence of specific heat and
$\phi$- dependence of thermal conductivity $\kappa_{zz}$ for various
fixed $\theta$. For planar field ($\theta$=90)
cusps appear when the field points along the node directions which
are equivalent to the [100] and [010] axis directions ($\phi$ =0,
$\pm$90$^\circ$) for the A- phase. In the B- phase only twofold
oscillations with cusps for field direction along [010] ($\phi$ =
$\pm$90$^\circ$) occur.

Very recently from the $\mu$SR experiment a possibility for triplet sc
in \PR is suggested \cite{Aoki03}. If this is indeed the case, we can
still reproduce the angular dependence of thermal conductivity as
observed for both A-phase and B-phase, using a gap function
$|\Delta(\vk)|\exp(\pm i\varphi)$ which has the same modulus as
eq.~(\ref{GAP2}) multiplied by a phase factor containing the azimuthal
angle $\varphi$ of \vk. This would then be a ``p+h'' wave gap instead
of the ``s+g'' wave gap in eq.~(\ref{GAP2}). Alternative triplet gap
functions have been proposed by \cite{Goryo02,Miyake03,Ichioka03}. On
the other hand, in this instance the impurity scattering is similar to
d- wave superconductors. This gives a dominant term in $\kappa_{zz}$
proportional to H$^\frac{3}{2}$, which is also different from
experiment \cite{Izawa02a}. In principle, field and temperature dependence and
impurity dependence of the angular part of magnetothermal conductivity
will allow to distinguish between singlet and triplet case. For this
purpose more detailed experiments for at lower temperatures ($<$ 0.2
K) are highly desirable.

\section{Concluding remarks}

We present here a simple nodal hybrid gap function model for \De which
appears to describe 
many features of the angular dependent magnetothermal conductivity in
the multiphase superconductor
PrOs$_4$Sb$_{12}$. From the thermal conductivity we infer the presence of point
nodes at \vk =(1,0,0), (0,1,0),(-1,0,0) and (0,-1,0) in A-phase and
at \vk =(0,1,0) and (0,-1,0) in B-phase. We have shown that the hybrid
nature of the sc order parameter leads to a very unusual effect of
impurities. Already for arbitrary small scattering $\Gamma$ an energy gap
opens immediately with related low temperature exponential behaviour
of specific heat and thermal conductivity. This implies the extreme
sensitivity of s+g- wave superconductivity to the presence of impurities.
We hope that the present work will stimulate further
experimental investigations in the muliphase superconductivity of skutterudite
PrOs$_4$Sb$_{12}$.\\

\acknowledgements

The authors thank J. Goryo for useful discussions.
K. Maki thanks the hospitality and support of Max-Planck-Institute for
the Physics of Complex Systems where part of this work was performed.
Q. Yuan acknowledges the partial support by the National Natural Science
Foundation of China (Grant No. 19904007).

\newpage

\end{document}